\documentclass[reprint,groupedaddress,longbibliography]{revtex4-1}

\usepackage{microtype}
\usepackage{graphicx}
\usepackage{amssymb}
\usepackage{wasysym}

\begin{document}

\title{Optical diffraction for measurements of nano-mechanical bending}

\author{Rodolfo I. Hermans}
\email[]{r.hermans@ucl.ac.uk}
\affiliation{London Centre for Nanotechnology, University College London, 17-19 Gordon Street, London WC1H 0AH, UK}
\affiliation{Department of Physics and Astronomy, University College London, Gower Street, London WC1E 6BT, UK}

\author{Benjamin Dueck}
\affiliation{London Centre for Nanotechnology, University College London, 17-19 Gordon Street, London WC1H 0AH, UK}
\affiliation{Leica Camera AG, Am Leitz-Park 5, 35578 Wetzlar, Germany}

\author{Joseph Wafula Ndieyira}
\affiliation{London Centre for Nanotechnology, University College London, 17-19 Gordon Street, London WC1H 0AH, UK}
\affiliation{Jomo Kenyatta University of Agriculture and Technology, Department of Chemistry, PO Box 62000, Nairobi, Kenya}

\author{Rachel A. McKendry}
\affiliation{London Centre for Nanotechnology, University College London, 17-19 Gordon Street, London WC1H 0AH, UK}

\author{Gabriel Aeppli}
\email[]{gabriel.aeppli@psi.ch}
\affiliation{London Centre for Nanotechnology, University College London, 17-19 Gordon Street, London WC1H 0AH, UK}
\affiliation{Department of Physics and Astronomy, University College London, Gower Street, London WC1E 6BT, UK}
\affiliation{Departments of Physics, ETH Z\"{u}rich, CH-8093 Z\"{u}rich, Switzerland and \'{E}cole Polytechnique F\'{e}d\'{e}rale de Lausanne (EPFL), CH-1015 Lausanne, Switzerland}
\affiliation{Synchrotron and Nanotechnology Department, Paul Scherrer Institute, CH-5232, Villigen, Switzerland}

\date{\today}

\begin{abstract} 
Micro-mechanical transducers such as cantilevers for atomic force microscopy often rely on optical readout methods that require the illumination of a specific region of the structure. Here we explore and exploit diffraction effects that have been previously neglected when modelling cantilever bending measurement techniques. The illumination of the cantilever end causes an asymmetric diffraction pattern at the photo-detector that significantly affects the calibration of the measured signal in the popular optical beam deflection technique (OBDT). The conditions for linear signals that avoid detection artefacts conflict with small numerical aperture illumination and narrow or smaller cantilevers. On the other hand, embracing diffraction patterns as a physical measurable allows a more potent detection technique that decouples tilt and curvature and simultaneously relaxes the requirements on the illumination alignment and detector position. We show analytical results, numerical simulations and physiologically relevant experimental data demonstrating the utility of the diffraction patterns. We offer experimental design guidelines and quantify possible sources of systematic error of up to 10\% in OBDT. We demonstrate a new nanometre resolution detection method that can replace OBDT, where Frauenhofer and Bragg diffraction effects from finite sized and patterned cantilevers are exploited, respectively. Such effects are readily generalized to arrays, and allow transmission detection of mechanical curvature, enabling instrumentation with simpler geometry. We highlight the comparative advantages over OBDT by detecting molecular activity of antibiotic Vancomycin, a representative example of possible multi-maker bio-assays.
\end{abstract}


\keywords{cantilever,AFM,calibration,far-field,diffraction,label-free,antibiotic,bio-assays}
\maketitle

\section{Introduction\label{s.intro}}

Micro-cantilevers are the most widely deployed micro-mechanical system (MEMS), initially developed for atomic force microscopy~\cite{Binnig1986}, but now serving as ultra-sensitive force transducers for applications ranging from airbag release to motion detection in mobile telephones. They have enabled nanobiotechnology~\cite{Muller2008,Dufrene2008}, branching beyond imaging into single-molecule manipulation and force metrology~\cite{Muller2008,Galera-Prat2012}, as well as multifunctional lab-on-a-tip~\cite{Muller2008} techniques. Cantilevers are promising for future medical diagnostic devices because they are both sensitive, with unlabeled biomolecules detected down to femtomolar concentrations within minutes~\cite{Rijal2007,Waggoner2009}, and because they can be multiplexed on arrays that allow multiple simultaneous differential measurements\cite{HuberF.2013,Ndieyira2008c}.
The biochemical sensitivity of cantilevers derives from the ability to detect small motions of their untethered ends, usually via the optical beam deflection technique (OBDT)~\cite{Meyer1988a,Alexander1989} implemented extensively for AFM-like devices. While conceptually simple, the need for careful alignment by specialists and a laser spot size small compared to the dimensions of the cantilevers limit general applicability outside of specialized research laboratories as well as the miniaturization needed both for enhanced sensitivity and massive multiplexing.
One reason for the preeminence of OBDT is that when the first atomic force microscopes were developed 30 years ago, inexpensive digital imaging (DI) was unavailable. In this paper, we describe how cheap DI enables a much more robust method, namely far field diffractive imaging, for optical readout of cantilever arrays. The method operates with light beams which can be much larger than individual cantilevers and whose angle of incidence and reflection need not be precisely set and measured, thus removing the obstacles presented by OBDT for non-expert use, miniaturization and multiplexing, and thus opening optically read- out cantilever arrays to numerous applications outside specialist research laboratories. It relies on the interference fringes easily visible for all objects with features on the scale of the wavelength of light, and we illustrate it for ordinary cantilevers, where the fringes are derived from their edges, as well as cantilevers into which we have inserted – using focused ion beam prototyping - periodic arrays of slots to create gratings whose diffraction patterns are very sensitive to bending.
The paper starts with a mathematical description of the interference effects associated with all optical readouts of cantilever bending, and then describe our tests of the diffractive method for unpatterned and patterned cantilevers, first for remote temperature sensing and then for biomedicine, where we examine antibiotic action.

\begin{figure}[tbh] 
 \begin{center}
  \includegraphics[width=0.98\columnwidth]{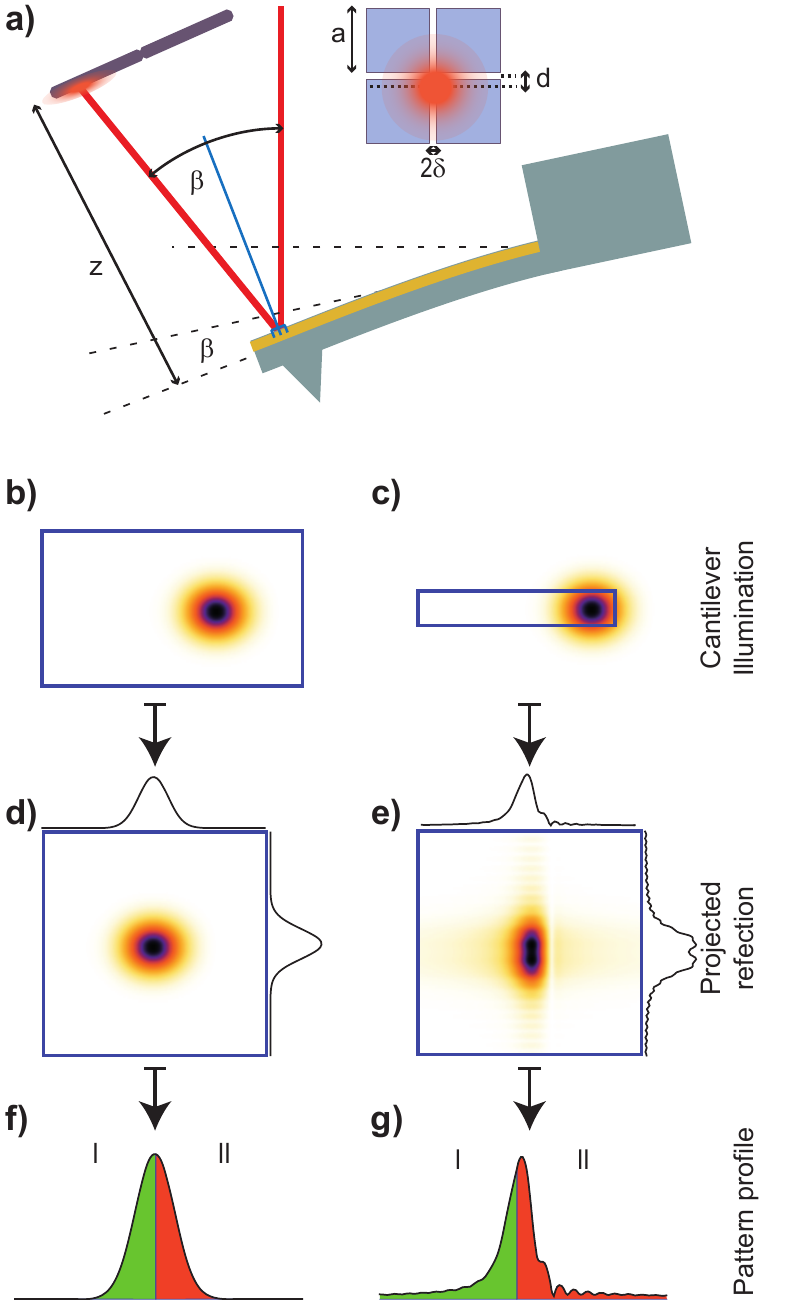}\\
  \caption{%
  \textbf{a}, Operation principle of the classical optical beam deflection technique (OBDT): Bending of the cantilever changes the reflection angle of a light beam. The reflected beam is projected onto a split photo-detector (also in inset) and the top-bottom differential signal is assumed proportional to the bending (See equation~(\ref{eq:gain})).%
  \textbf{b}, Ideal case, a cantilever is illuminated by a beam much smaller than its width, causing a Gaussian reflected beam projected on the split photo-detector (\textbf{d}).%
  \textbf{f}, The projected profile is symmetric as is therefore also the sensitivity curve.%
  \textbf{c}, Non-ideal case, the cantilever edges are illuminated causing %
  \textbf{e}, a cross-shaped diffraction patter projected onto the spit photo-detector.%
  \textbf{g}, The projected profile as well as the sensitivity curve are asymmetric.}
  \label{fig:OBDT-Diffraction}
 \end{center}
\end{figure}

\section{Huygens-Fresnel description\label{s.HF}}

Optical techniques for MEMS metrology require the illumination of a region or all of the device probed. We focus our attention on the details of OBDT depicted in Figure~\ref{fig:OBDT-Diffraction}a, whose operation consists of reflecting a focused or collimated light beam from the free end of a cantilever and measuring the position of the reflection projected onto a segmented photo-diode or similar position-sensitive device.

We model the optical system using the Huygens-Fresnel principle, where the light re-emitted by the cantilever, both reflected or transmitted, can be understood as the summation of an infinite number of infinitesimal point sources located on the cantilever surface. A cantilever acts as a rectangular slit source in the plane $(\xi, \eta)$ and under Fresnel's approximation the optical wave in the observing plane $(x,y)$ is given by~\cite{goodman2005introduction}

\begin{equation}\label{eqn:fresnel-conv}
    U(x, y) = \frac{e^{i k z}}{i \lambda z} \iint\limits_{-\infty}^{\infty}  U(\xi, \eta, z)\, e^{ i \frac{\pi}{\lambda z} \left[ (x-\xi)^2+(y-\eta)^2 \right]} d\xi d\eta.
\end{equation}

where $U(\xi, \eta, z)$ is the function defining the amplitude and phase at the re-emitting source, defined by the illuminating profile and the geometry of the device.

The beam projected onto the device is typically assumed to be radially symmetric around its maximum, in our case we assume a 2D Gaussian illumination centred in the origin,   $U(\xi, \eta, z)= \exp(-\frac{1}{2} \frac{\xi^2 + \eta^2}{\sigma^2 })$.

\subsection{Ideal infinite plane}

In the ideal case of a perfectly flat, infinite reflecting surface, the incident beam will be reflected unperturbed into a detector, maintaining the original profile. We model the finite size of the detector and consider the Gaussian beam projected on a \mbox{4-segment} photodiode and calculate that the differential signal of the segmented sensor versus beam displacement is (See supplementary material for derivation of exact solution, approximations, analysis, and figures)

\begin{equation}\label{eq:gain}
V(d)\approx G    \frac{d}{\sigma} \, \, \text{\textrm{Erf}}\left(\frac{a^2}{ 2 \sigma ^2}\right)\,\left[ e^{-\frac{\delta ^2}{2 \sigma ^2}}\,\, \left(1-e^{-\frac{a^2}{2   \sigma ^2}}\right)\right]
\end{equation}

where the factor $G$ is proportional to the laser intensity and the electronic gain, $d$ is the distance of the centre of the beam from the centre of the detector, $\sigma$ the standard deviation of the Gaussian beam at the detector, \textrm{Erf} is the error function, $a$ is the length of the photodiode segments and $2\delta$ is the distance between these segents (see figure 1 for graphical definitions of $\delta$, $a$ and $d$). The approximation $V(d) \approx G d/\sigma$ implies maximum gain and linearity; and it is valid within 1\% only if $7 \delta < \sigma < a/3 $, and $ |d| < \sigma/4 $ (see supplementary material). We observe in equation~(\ref{eq:gain}) that a smaller spot size $\sigma$ appears to increase the gain, as previously reported~\cite{DCosta1995,Butt20051}, but it must be noted that this is true only if $\delta$ is much smaller than $\sigma$. Because the signal is linear within 1\% only for  $|d| < \sigma/4 $ a small spot size will also restrict the dynamic range.

For beams not collimated but focused at a finite distance, $\sigma= z \theta$ and $d = 2 z \beta$ (where $z$ is the detector distance, $\beta$ is the cantilever deflection angle, and $\theta$ the beam divergence) and therefore the signal $G \beta/\theta$ is independent of $z$ and is maximized by reducing the beam divergence $\theta$. Considering that the minimum beam cross-section diameter is given by  \mbox{$\phi = {4\lambda}/{\pi\, \theta}$}, we see that a small beam divergence determines the minimum size of the laser spot and consequently the minimum width of the cantilever, if diffraction is to be avoided, as we will see.

\subsection{Implications for standard readout }

The infinite plane model could only be valid if the reflected beam cross-section is completely contained in a flat reflecting surface (Figure~\ref{fig:OBDT-Diffraction}b), otherwise any illuminated edge of the cantilever (Figure~\ref{fig:OBDT-Diffraction}c) will cause a diffraction pattern to appear on the photo-detector.

To model the influence of a finite cantilever we first separate variables using $U(x,y) = -i \exp(i k z)\, \mathcal{I}(x) \mathcal{I}(y)$  and concentrate attention on the longitudinal axes $x$,$\xi$.
We now introduce an edge at position $\xi=s$, and re-define $U(\xi)=I'\,H(\xi-s) \, \exp(-\frac{1}{2} (\frac{\xi }{\sigma })^2)$, where $H(x)$ is the Heaviside step function, $I'= I_o/\sigma \sqrt{2 \pi}$ to obtain:

\begin{eqnarray}\label{eq:Fresnel1}
\mathcal{I}(x) &=& \frac{I'}{\sqrt{\text{$\lambda $z}}}  \int\limits_{-\infty}^s    e^{\frac{\pi i }{\lambda z} (x-\xi )^2} \, e^{-\frac{1}{2} \left(\frac{\xi }{\sigma }\right)^2} \, d\xi\, ,\label{eq:Fresnel}
\end{eqnarray}

where $\xi$ is the coordinate along the cantilever, $x$ along the detector, $\lambda$ is the illumination wavelength, $s$ the position of the cantilever edge and $\sigma$ the standard deviation of the Gaussian beam.

\begin{figure}[b] 
 \begin{center}
  \includegraphics[width=\columnwidth]{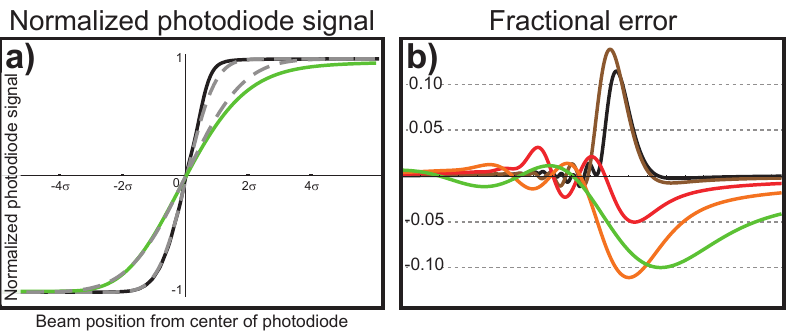}\\
  \caption{%
  \textbf{a}, Normalized differential photodiode signal versus beam position (cantilever bending) for an infinite detector at different observation distances, in solid lines considering the diffraction caused by the cantilever edge and in dashed lines neglecting diffraction. %
  \textbf{b}, Magnitude of the fractional error as a function of bending for different conditions \mbox{$\lambda z/\sigma^2$ = 1,2,5,10} and 20 in colours black, brown, red, orange and green respectively. The error as a consequence of neglecting the diffraction phenomena can surpass 10\% in some cases.}
  \label{fig:SignalCurveAndResidue}
 \end{center}
\end{figure}

Figure~\ref{fig:OBDT-Diffraction}d and~\ref{fig:OBDT-Diffraction}e shows calculated diffraction patterns caused by a Gaussian beam reflected from cantilevers with edges respectively far and close to the beam centre. Illuminating the cantilever's edge causes a broad-tailed asymmetric diffraction pattern, significantly different from the typically assumed Gaussian intensity distribution. This diffraction artifact causes an asymmetric dependence of the measured signal on the cantilever deflection, as shown in figure~\ref{fig:SignalCurveAndResidue}a. The artifact is negligible for controlled systems, where a feed-back loop maintains the cantilever bending at a small constant value and the excursions from this value are small during experimentation. Interestingly, the asymmetry of the sensitivity could become relevant for uncontrolled systems, such as bio-markers, and systems where calibration and measurement happens at opposite sides of the deflection curve. For instance in single-molecule force-spectroscopy experiments where signal versus cantilever bending calibration curves are obtained from upward-bending (pushing the cantilever into the surface) long trajectories whereas sample measurements are downward-bending for a pulling experiment.\cite{Galera-Prat2012,Hutter1993}  Figure~\ref{fig:SignalCurveAndResidue}b shows the normalized difference between considering or neglecting diffraction, implying that bending could be under or overestimated by more than 10\%.

We have seen that the ideal case for OBDT is the limit when the illuminating beam is much smaller than the cantilever. In practice, small numerical aperture systems create focal spots comparable in size to cantilever widths ($10-100\mu m$). Carefully aligning and focusing a small laser spot onto the centre of a cantilever end, to avoid the diffraction caused by the edges, can become a tedious task, if at all possible. We investigate the opposite limit, when the illuminated area is much bigger than the cantilever, and the diffraction pattern caused by the finite cantilever size contain all the information we need.

\section{Cantilever diffraction detection method}

\begin{figure}[t] 
 \begin{center}
  \includegraphics[width=\columnwidth]{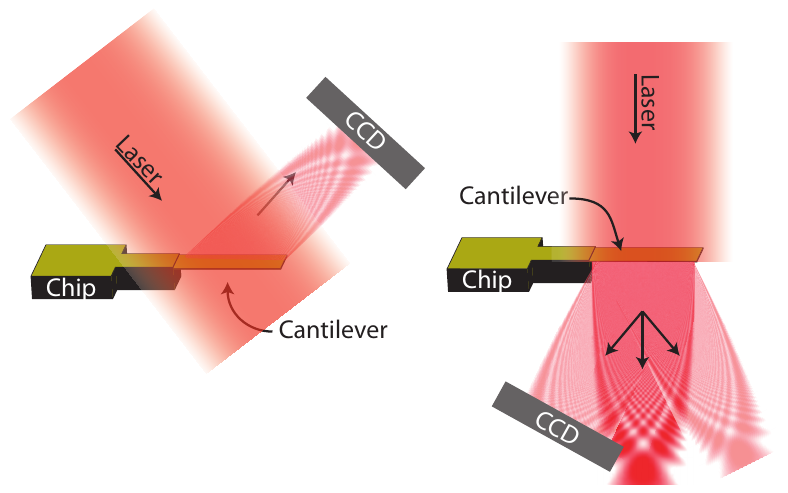}\\
  \caption{%
  Two possible diffraction read-out operation modes. On the left the light reflected from the finite size cantilever causes a diffraction pattern. On the right the light transmitted through the cantilever causes the diffraction pattern. Even though typical transmittance of solid cantilevers could be insufficient to provide an acceptable signal to noise ratio in transmission mode, the fact that the features of interest in the diffraction pattern \emph{i.e.} position shift and magnification  are independent of cantilever shape, any arbitrary pattern of holes through the cantilever would allow higher transmittance. In particular an array of slits provides the advantage of high intensity high order diffraction peaks detectable at wide angles away from the direct incident beam.}
  \label{fig:Method}
 \end{center}
\end{figure}

Now we demonstrate that the shape of the diffraction pattern reveals the details of a surface curvature. Figure~\ref{fig:Method} shows a cartoon of two operational modes of the new proposed readout method. A broad laser source illuminates the whole cantilever with close to homogeneous intensity while a CCD or CMOS detector captures some of the diffraction fringes created by either reflected or transmitted light~\cite{AEPPLI2008}.

We have previously shown that we can describe a rectangular cantilever of dimensions $(w,l)$ curved along the $\xi$ axis by defining~\cite{Hermans2013}

\begin{equation}\label{}
  U(\xi, \eta) = S_{\xi} S_{\eta}  \, \exp \left[\frac{4 \pi i}{\lambda}\,  \left( a \xi+ b \xi^2\right) \right];
\end{equation}

where $S_{\xi}=\text{rect}(\xi/l)$, $ S_{\eta}=\text{rect}(\eta/w)$ define the dimensions of the cantilever, $a$ and $b$ are the coefficients of the quadratic shape that describes the cantilever tilt and curvature and the exponential in the right-hand-side models the phase from the difference in  optical path caused by the bending of the surface.  By rearranging equation~(\ref{eq:Fresnel1}) for a perfect square binomial we can write

\begin{equation}\label{eq:AmplFresnelPattern}
\mathcal{I}(x) = A  \int\limits_{-\infty}^{\infty} S_{\xi} \exp \left[ i \frac{\pi}{\lambda z} \left(\frac{x-2az}{1+4bz}-\xi \right)^2 \right] d\xi ;
\end{equation}

where $ A =  (\lambda z)^{-\frac{1}{2}}  \exp\left[ i \frac{4 \pi}{\lambda \, (1+4 b z)}  \left(b x^2 + a x - a^2 z \right) \right]$. We define

 \begin{equation}\label{eq:changevar}
    x'= \frac{x-2az}{1+4bz}
 \end{equation}

 and observe that the term $2az$ causes a shift and $1+4bz$ a magnification of the diffraction pattern.\cite{Hermans2013} Applying the condition $z\gg \xi^2/\lambda$  the Fraunhofer approximation for the far field $\exp(2\pi i \, \xi^2 /\lambda z)\approx 1$ implies

\begin{eqnarray}\label{eq:AmplFraunhoferPattern}
\mathcal{I}(x) &\approx &A e^{i \frac{\pi}{\lambda z} x'^2} \int\limits_{-\infty}^{\infty} S_{\xi}\, \exp \left[ i \frac{\pi x'}{\lambda z}  \xi \right] d\xi, 
\end{eqnarray}

and therefore the observed intensity profile is given by

\begin{eqnarray}\label{eq:IntFraunhoferPattern}
|\mathcal{I}(x)|^2 & \approx & \frac{1}{\lambda z}   \left|\, \int\limits_{-\infty}^{\infty} S_{\xi}\, \exp \left[ 2 \pi i \left(\frac{x'}{2 \lambda z}\right)  \xi \right] d\xi \right|^2.
\end{eqnarray}

 We see that also in the Fraunhofer far-field approximation the diffraction pattern caused by a micro-structure experiences a magnification given by the curvature of the surface and a shift in position given by the tilt of the surface as defined by the change of variables in equation~(\ref{eq:changevar}). Therefore, tracking the changes in position and shape of a diffraction pattern enables to monitor the tilt and curvature of the cantilever independently. This result holds for different cantilever shapes $S_{\xi}$ in reflection mode.

 To model the transmission mode, we took a step back and considered the Fresnel approximation not in a plane but from a curved source. According to the Huygens-Fresnel principle the field intensity is given by~\cite{goodman2005introduction}

 \begin{equation}
U(x,y)= \iiint \frac{z}{i \lambda r^2}  \,  U(\xi ,\eta, \zeta ) \, \exp{\left( \frac{2 \pi i }{\lambda } r \right)} d\xi d\eta d\zeta
 \end{equation}

 where $r=\sqrt{(x-\xi )^2+(y-\eta )^2+(z-\zeta)^2}$ is the distance from the virtual source point at $(\xi,\eta,\zeta)$ and the observation at $(x,y,z)$. We consider a source curved on $\xi$,  $U(\xi, \eta, \zeta) = S_{\xi} S_{\eta} \delta (\zeta- a \xi - b \xi^2 )$ and follow similar procedure as before but with $z$ replaced by $z +  a \xi + b \xi^2$. We approximate $r$ up to second order in the numerator exponent and re-arrange to complete squares

 \begin{equation}
   r \approx A_{\text{Tr}} + \, \frac{m}{2 z} \left(\xi -\frac{x-n}{m}\right)^2
 \end{equation}
where the zero order term $A_{\text{Tr}}$ will later cancel with its complex conjugate when calculating the intensity and

\begin{eqnarray}
  n & = &a z - a\frac{x^2}{2 z} \\
  m &=  &1 + 2 b z + 2 a \frac{x}{z} -b \frac{x^2}{z}
\end{eqnarray}

To the extent that $x\ll z$ we can neglect the terms containing $x$ in the right hand side and recover a similar result as before, this time only for small changes in cantilever tilt and curvature, implying a pattern shift of $az$, pattern magnification of $1+2 b x$ respectively. These magnitudes differ from the previous result by a factor of two because, for a given cantilever displacement, light travels the path only once in transmission mode but twice in reflection mode.

We next explore experimentally both the case of a flat cantilever and also the case where the cantilever has a series of narrow slits forming a diffraction grating.


\begin{figure}[tb] 
 \begin{center}
  \includegraphics[width=\columnwidth]{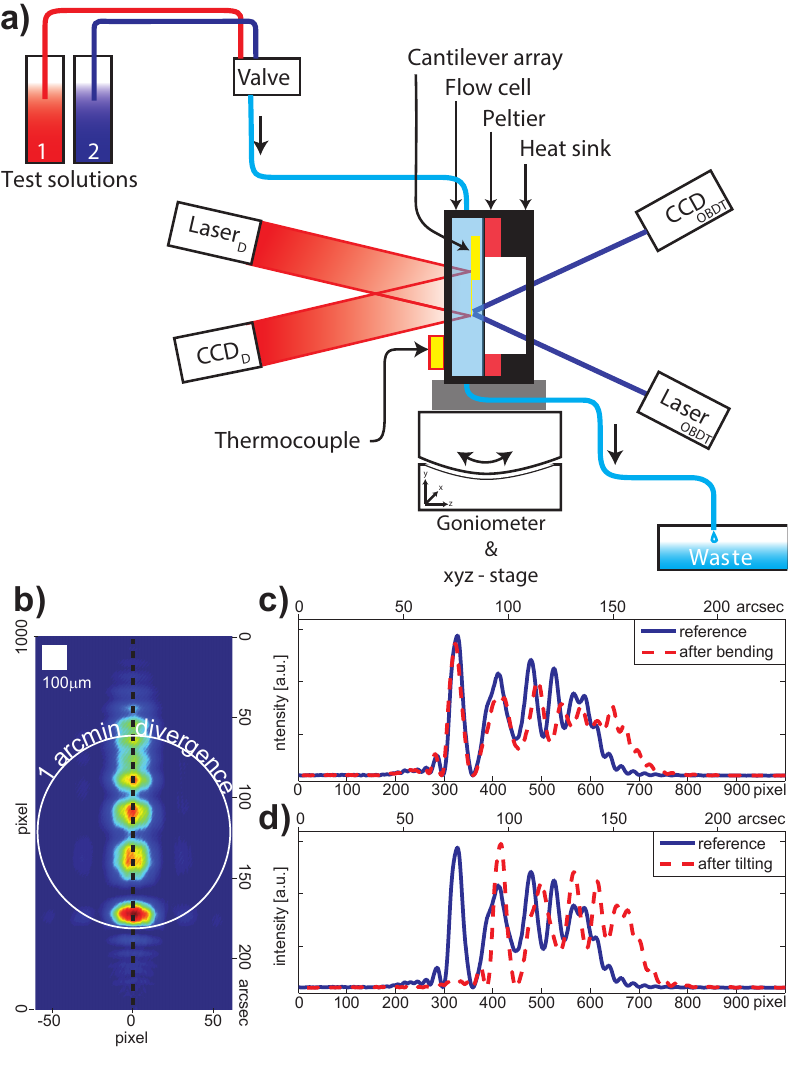}\\
  \caption{%
  \textbf{a},~Schematic design of the experiment. The cantilever is mounted vertically in the flow cell. Test solutions 1 and 2 are selected by a valve and driven through the cell into the waste by gravity flow. %
  On the right side, the optical beam deflection technique (OBDT) is used to read out the cantilever bending and %
  on the left side is the new diffractive readout in reflection mode consisting in broad beam HeNe Laser illumination (632.8 nm, 5.0 mW HRR050 Thorlabs) and CCD detection. Goniometer (RV160CCHL, VP-25X from Newport), CCD are controlled by one LabView program. The TCM controller for the Peltier module and the pico-logger for the external thermocouple were controlled by two separate programs.  %
  \textbf{b},~Reflection-mode 2D diffraction pattern from a flat cantilever captured by a CCD ORCA-AG Hamamatsu %
  \textbf{c},~Comparison of the cross-section of reference and measured pattern after a change in curvature (Temperature change from 25$^\circ$C to 24$^\circ$C) and %
  \textbf{d},~after tilting  approximately 25 arcsec. %
  }
  \label{fig:experiment}
 \end{center}
\end{figure}

To verify the usefulness and performance of this detection method we have built the trial setup of Figure~\ref{fig:experiment}a. We used a cantilever array windowed flow cell that allowed simultaneous measurements using the new diffractive readout method and the classic OBDT. We capture diffraction patterns generated by the cantilevers and study the changes as the cantilever rotates (goniometer tilt) and curves (temperature change).  Figure~\ref{fig:experiment}b shows the 2D pattern acquired with a CCD.  Figure~\ref{fig:experiment}c-d shows the pattern profile and confirms experimentally our prediction that, independent of the details of the pattern, changes in curvature magnify the pattern profile and changes in tilt only displaces the pattern without significant deformation.

\begin{figure}[tb] 
 \begin{center}
  \includegraphics[width=\columnwidth]{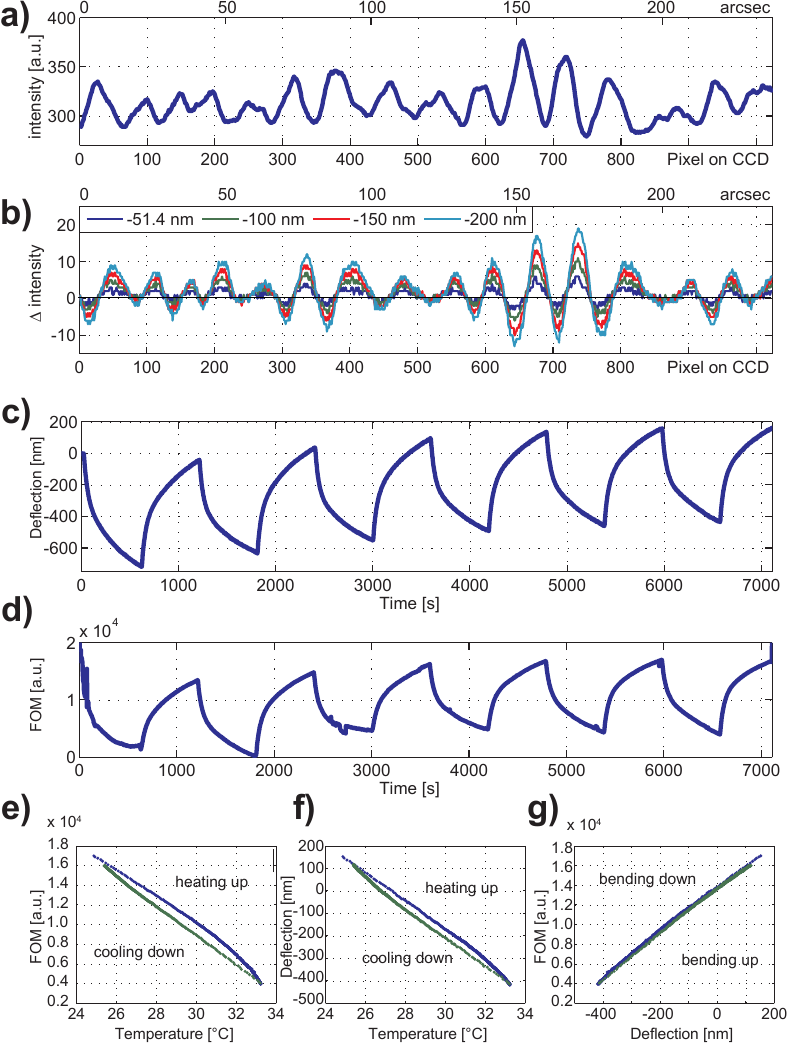}\\
  \caption{%
  Reflection measurements with non-patterned cantilevers. %
  \textbf{a}, Cross-section intensity of a diffraction pattern is used as a reference.%
  \textbf{b}, Four representative differences between the observed and the reference pattern is calculated for cantilever deflections during a temperature cycle (inset distances via OBDT correspond to changes in temperature of approximately $0.7^{\circ}C$,$1.4^{\circ}C$,$2.0^{\circ}C$ and $2.8^{\circ}C$ respectively).%
  \textbf{c}, Cantilever deflection measured by OBDT as a function of time while the cantilever is cycled in temperature from $25^{\circ}C$ to $33^{\circ}C$. %
  \textbf{d-g}, The figure of merit (FOM) calculated as the root mean squared value of the pattern differential features an excellent correlation with the deflection measured with OBDT.%
  }
  \label{fig:bending}
 \end{center}
\end{figure}

To exemplify the data acquisition procedure Figure~\ref{fig:bending} shows measurements for the cantilever as the temperature is cycled in the range 25$^{\circ}$C to 33$^{\circ}$C. The gold-coated silicon cantilever acts as a bimetallic strip and the differential thermal expansion causes a homogeneous curvature of significant magnitude~\cite{Thundat1994a}. The observed deflection is around 72~nm/$^{\circ}$C. A reference diffraction pattern is recorded by the CCD camera at the beginning of the experiment (Figure~\ref{fig:bending}a) as a spatial array of intensities. At the same time the initial position of the OBDT spot in the CCD is recorded also as a reference. All further patterns and spot positions are measured sequentially in time and compared with their respective references. The difference between the pattern intensity arrays are calculated (Figure~\ref{fig:bending}b). We define a figure of merit (FOM) as the root mean square value of the difference between the observed pattern and the reference pattern (Figure~\ref{fig:bending}b).

\begin{equation}
  \text{FOM}= \sqrt{\sum_{(i,j)} (I_{(i,j)}-R_{(i,j)})^2}
\end{equation}

\mbox{Figure~\ref{fig:bending}g} shows that the FOM calculated from the diffraction pattern closely correlates with the measurements from OBDT evidencing that the far-field diffraction readout can replace the classic OBDT.

\begin{figure}[tb!] 
 \begin{center}
  \includegraphics[width=\columnwidth]{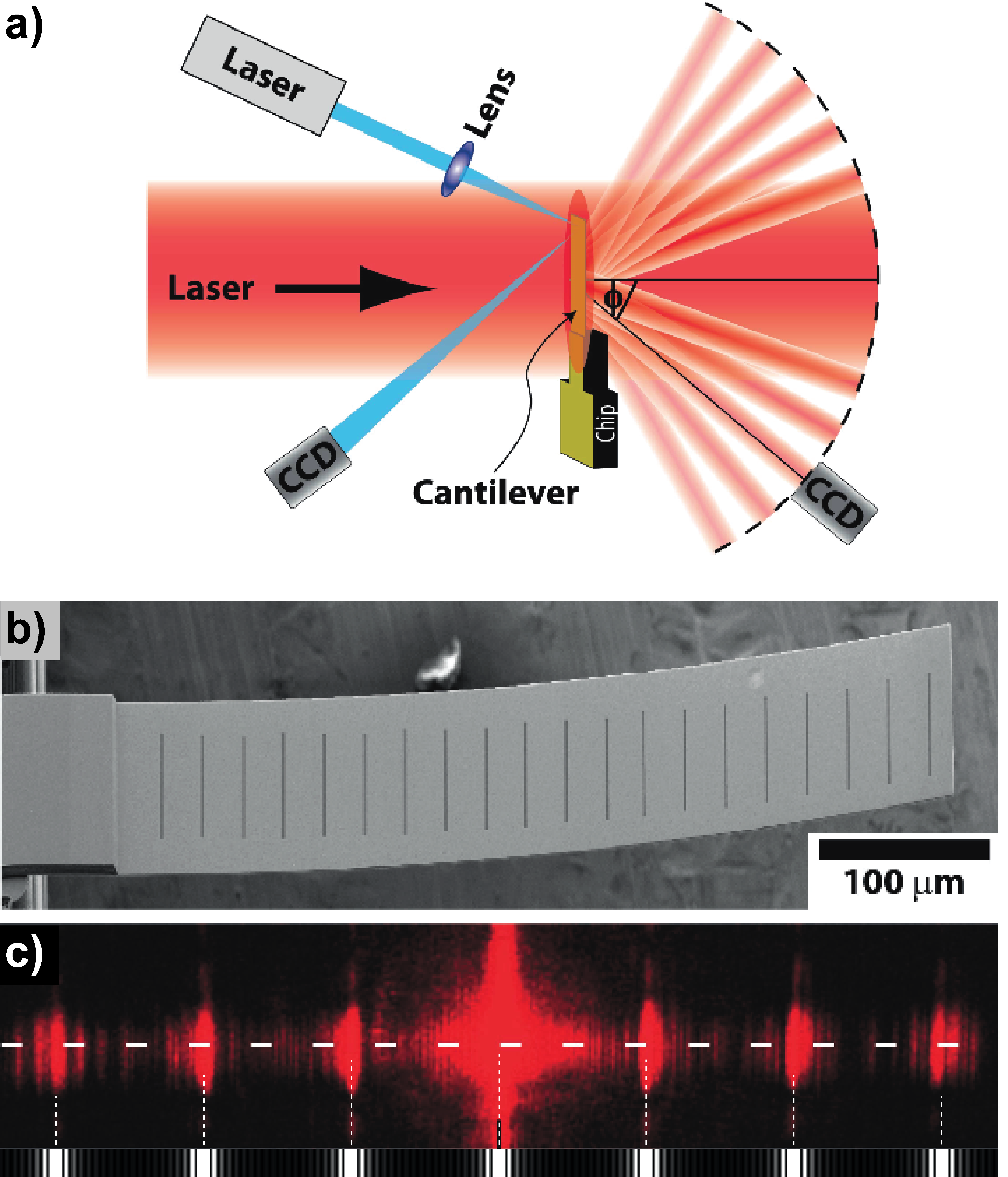}\\
   \caption{%
   \textbf{a}, Transmission mode experimental configuration: a broad laser beam illuminates the whole cantilever for diffraction read-out while simultaneously a narrow focused laser illuminates the cantilever tip for optical beam deflection technique (OBDT).%
   \textbf{b}, A cantilever featuring an array of slits created by Focused Ion Beam (width $w=1 \mu m$, spacing $s= 23.4 \mu m$) can act as diffraction grating allowing detection using high order Bragg peaks. %
   \textbf{c}, Measurement of a series of Bragg (strong red) and Fraunhofer (weak red) fringes created by the illuminated slit array compared with predicted fringes (white) by the diffraction model for a straight cantilever $\text{sinc}(\beta )^2 \left(\frac{\sin (\alpha  n)}{\sin (\alpha )}\right)^2$ with $\beta = \pi w/\lambda  \sin (\theta )$ ,  $\alpha = \pi  s/\lambda \sin (\theta)$, $n=20$ and $\lambda=632.8 nm$.
   }
  \label{fig:experiment2}
 \end{center}
\end{figure}

\begin{figure}[tb!] 
 \begin{center}
  \includegraphics[width=\columnwidth]{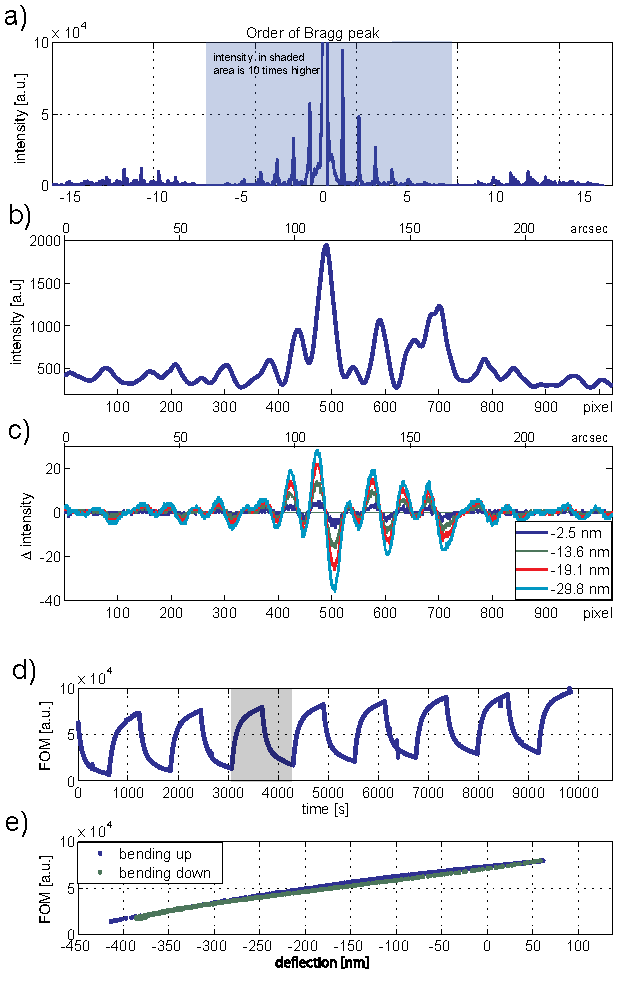}\\
  \caption{Transmission measurements with patterned cantilevers.%
   \textbf{a}, Diffraction pattern from -15th to +15th order. The intensity plotted in the shaded area has been displayed reduced by a factor 10 to increase the visibility of higher order Bragg peaks. %
   \textbf{b}, Initial diffraction pattern showing the 19th order Bragg peak and subsidiary peaks. %
   \textbf{c}, The difference of diffraction patterns to the initial pattern where the bending is relative to the initial bending. %
   \textbf{d}, The response of the cantilever by cycling the temperature by approximately 5$^\circ C$. Figure of merit (FOM) computed from diffraction pattern. %
   \textbf{e}, FOM versus deflection as measured with OBDT.%
   }\label{fig:ExpTrans}
 \end{center}
\end{figure}

\begin{figure}[tb!] 
 \begin{center}
  \includegraphics[width=\columnwidth]{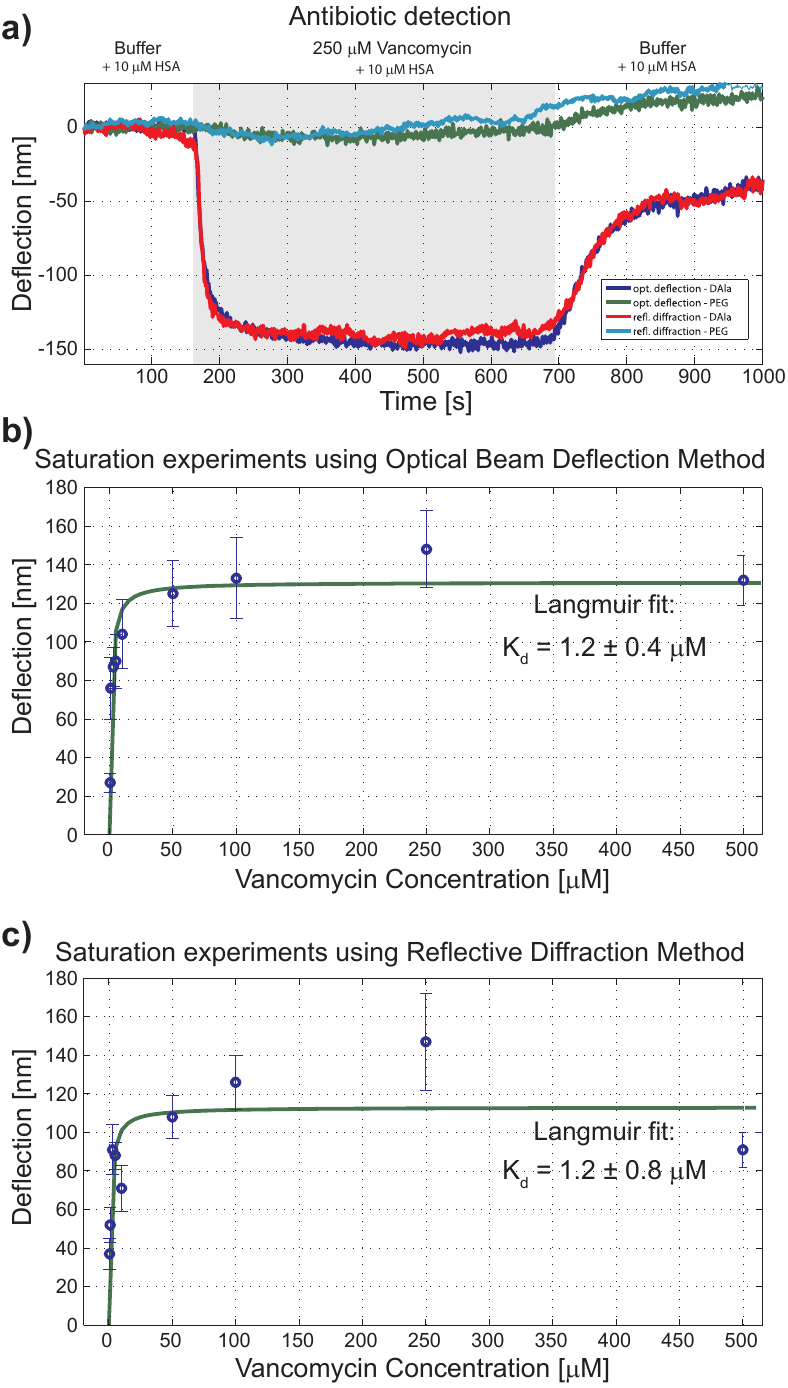}\\
  \caption{%
  Experimental estimation of dissociation constant of Vancomycin via saturation experiments in cantilevers is performed simultaneously with both techniques, reflective diffraction and OBDT.
  \textbf{a}, The dramatic bending observed on the functionalized cantilever is attributed to the surface pressure caused by the specific ligand-receptor binding of Vancomycin and a bio-mimetic bacterial cell wall target. In contrast, passivated cantilevers feature negligible bending. Equal equilibrium bending is estimated by both by OBDT and diffraction detection techniques. The equilibrium deflection measured at different concentrations of Vancomycin are fitted by a Langmuir isotherm function. Identical values of the dissociation constant $K_d$ are obtained using \textbf{b} OBDT and \textbf{c} our diffraction technique. Data points are mean values and error bars standard deviation, with $n=3$.}
  \label{fig:VancoTrace}
 \end{center}
\end{figure}

Previous results can be reproduced both in reflection and transmission mode, but the latter may suffer from a poor signal to noise ratio if the cantilever features a small transmissivity. An interesting consequence of our analysis is that, where the approximations hold, the principle of diffraction readout applies independent of the form $S_{\xi}$ of the cantilever, particularly it applies also to a periodic array of slits. We performed a second experiment with a cantilever featuring a series of narrow slits created by the Focused Ion Beam technique as shown in Figure~\ref{fig:experiment2}b. Here the light transmitted through the array of slits causes an intense series of Bragg peaks (Figure~\ref{fig:experiment2}c, Figure~\ref{fig:ExpTrans}a) with spacing reciprocal in relation to the spacing of the slits. The weaker Fraunhofer peaks between the Bragg peaks result from the finite number of slits. More slits will increase the number of Fraunhofer fringes and decrease their intensity. The entire feature-rich pattern is sensitive to the phase differences caused by cantilever deflection and consequently, contrary to other techniques, a patterned cantilever allows the detection of bending across a broad range of detection angles.

To further test the method in challenging conditions of practical interest we reproduce previous results on binding of antibiotics to target peptides.\cite{Ndieyira2008c}
Un-patterned cantilevers were either sensitized or passivated by selectively forming a self-assembled molecular monolayer by incubating them individually in micro-capillary tubes. Passive control cantilevers were coated with polyethylene glycol (PEG) and target cantilevers were coated with drug-sensitive mucopeptide analogue in a procedure detailed elsewhere~\cite{Ndieyira2008c}. Figure~\ref{fig:VancoTrace}a shows bending of cantilevers coated with a bio-mimetic bacterial cell wall target in response to 250$\mu$M Vancomycin detected with both the diffractive method and OBDT. Upon repetition at different Vancomycin concentrations we obtain the saturation curves in Figures~\ref{fig:VancoTrace}b-c featuring identical dissociation constants, irrespective of the detection method.

\section{Advantages perspective}

Diffraction features generated by a microstructure such as a cantilever are exquisitely sensitive to geometrical details such as curvature, tilt, position of the edges and roughness of the surface.\cite{Hermans2013} We have shown that this sensitivity, depending on details, can on one hand yield artefacts that skew assumed calibrations of OBDT. On the other hand, they  provide alternative means for detecting independently changes of tilt or curvature. Avoiding diffraction from the cantilever surface requires comparatively narrow illuminating beams or broad flat reflection areas. If it exists at all, the optimum position and focus of the illuminating beam will tend to be narrow, and therefore continuous and tedious re-alignment and re-calibration could be necessary. Another common problem of highly focused laser beams in liquid is that the measured intensity becomes sensitive to transient perturbations caused by aggregates or other impurities crossing the beam at the narrow focus, be they suspended in the liquid or diffusing in the cantilever surface. There are several advantages to be gained by measuring the details of diffraction patterns, instead of trying to avoid them.

Keeping the laser spot larger than the cantilever dimensions makes exact knowledge of the laser spot position\cite{0957-4484-15-9-039} superfluous and therefore alignment becomes a fast, simple and reliable procedure. A broad illumination beam also minimizes the relative magnitude of perturbations caused by  particles crossing the illumination beam and eliminates temperature gradients in the cantilever~\cite{Thundat1994a,Chigullapalli2012}. The optical diffractive readout does not necessarily rely on having a reflective surface and therefore the choice of surface coatings is widened. Beside the reflection component, cantilevers featuring slit arrays allow high signal-to-noise levels in transmission mode, and more interestingly, the detector can be located off-axis around high order Bragg peaks, offering a much broader set of geometrical configurations for the detector. The relative change in size of the diffraction pattern is of particular interest because it is invariant to lateral translation and rotation and independent of the shift caused by changes in tilt, making the measurement intrinsically robust to small perturbations in the detector and cantilever positions and orientations.

We have provided an exact analytical model for parabolic bending and reflection mode and an approximate analytical solution for transmission mode when $x\ll z$. We have also defined a figure of merit (FOM) that allows an effective implementation of the detection technique independent of these analytical considerations or other modelling.


The high order Bragg fine structure resembles that exploited for oversampled X-ray crystallography\cite{Miao2000}. We have a visible light analog of the X-ray experiments and here the information from the phase difference created by the cantilever curvature is contained in the details of the intensity between Bragg peaks.

Measuring the details of a diffraction pattern requires a more complex detection device such as a CCD or CMOS sensor array, as opposed to a simpler split photodiode. This increase in complexity is justified by the increased amount of information available, as tilt and bending could be decoupled. CCD and CMOS sensor also feature reduced bandwidth, but this is not a limitation for probing systems with relevant time scales much longer than the CCD frame acquisition period, such as the ones shown here. It is also worth noting that the ubiquity of digital imaging today, especially as compared to when OBDT was developed in the early 1990s, makes the use of position-sensitive optical detection a very competitive option for modern low-cost instrumentation.

The presented far-field technique distinguishes itself from the related NANOBE~\cite{Hermans2013} in that it does not demand a lens to maintain the near field condition and allows a transmission mode at wide angles. At the far field, if more than one cantilever is illuminated at a time, the observed diffraction pattern will be sensitive to the differential displacement, while in the near field there is negligible overlap from the information from near cantilevers.


In summary, we have analysed mathematically the popular optical beam deflection technique (OBDT) for measuring cantilever deflection and found that the conditions for maximum gain,  linearity and symmetry require illumination spot sizes that are heavily constrained by the geometry of the detector and the cantilever, desired dynamic range and gain,  and the divergence of the illuminating beam. Ignoring such constraints can cause detection errors in exceed of 10\%.
We propose a diffraction readout method which decouples as independent observables the cantilever tilt and curvature of the cantilevers and does not require precise alignment. The excellent correlation observed (Figure~\ref{fig:bending}g) between the OBDT and our proposed diffraction method demonstrates the advantages of replacing OBDT with the more robust diffraction method. We have demonstrated the fundamental principles and practicality of our approach both analytically and experimentally for a clinically relevant application.

\section{Methods}

\begin{small}
\paragraph{Readout} We used a cantilever array chip (IBM) where each cantilever was $500\mu$m long, $100\mu$m wide and $0.9\mu$m thick, and coated with a layer of 2~nm titanium followed by 20~nm of gold. The array chip was mounted in an aluminium flow cell with sapphire windows at both sides and a thermoelectric Peltier element and thermocouple for temperature control. A broad laser beam (HeNe 632.8 nm, 5.0 mW, HRR050 Thorlabs)) illuminates the surface of a single  cantilever to test the diffractive readout method. Simultaneously, a narrowly focused laser beam was used to measure the cantilever bending using OBDT as control. Both reflected beams were projected onto CCD sensors (ORCA-AG from Hamamatsu, Pixel size: 6.45 $\mu$m $\times$ 6.45 $\mu$m and FireWire 400 Color Industrial Camera DFK 31AF03 with sensor Sony ICX204AK, Pixel size 4.65 $\mu$m $\times$ 4.65 $\mu$m) mounted on calibrated goniometers (Rotation Stage RV160CCHL and xyz-stage VP-25X from Newport) to allow recording the intensity at different angles. CCD sensors were approximately at a distance of 100~mm for reflection mode and 250~mm for transmission mode. Our raw data consist of the diffraction patterns generated by the cantilevers captured as 12bit TIFF images. Exposure times on the order of milliseconds were adjusted to avoid saturation and maximize dynamic range. The expansion of the pattern by approximately 12.5\% observed in Figure~\ref{fig:experiment}c corresponds to a change in curvature of $\delta b= 0.125/(4 z)\approx 0.3125~m^{-1}$  and a displacement at the end of the cantilever of $\delta b~(500\mu m)^2\approx78~nm$.

\end{small}
\newpage

%


\begin{acknowledgments}
G.A. and R.H. thank support from grant EPSRC EP/G062064/1, ``Multi-marker Nanosensors for HIV'', May 2009.
\end{acknowledgments}

\section{Competing financial interests}
R.I.H., B.D and G.A. are co-authors of two patents (US2010/0149545 and EP2156441 A1: G. Aeppli and B. Dueck: “Apparatus and Methods for measuring deformation of a cantilever using interferometry”, publication 24th Feb 2010 and WO 2013144646 A3: R. Hermans and G. Aeppli: “Measuring Surface Curvature”, publication 21st November 2013) whose value could increase if the methods and ideas described in this paper find widespread application.

\section{Author contribution}

R.I.H. developed the idea of the illuminated edge artefacts, the 4-segments photodiode constrains and the diffraction detection method analytical description.
B.D. J.N. R.A.M. and G.A. conceived the original idea of the far field cantilever diffraction detection methods.
B.D., J.N., and R.A.M. conceived the original ideas relating antibiotic detection experiments.
B.D. performed the FIB on cantilever to create the slits pattern.
B.D. and J.N. functionalized cantilever arrays and performed experimental measurements.
R.I.H. and B.D. analyzed data, interpreted results, designed the figures and wrote the manuscript.
J.N., R.A.M. and G.A. supervised the experiments and analysis.
All authors discussed the results and commented on the manuscript.
R.H. and B.D. contribute equally to this research.

\section{Additional information}
Supplementary information is available in the online version of the paper. Reprints and
permissions information is available online at www.nature.com/reprints. Correspondence and
requests for materials should be addressed to R.I.H. or G.A.


\end{document}